
%
%
\input harvmac
\def\cft{conformal field theory}
\def\ie{{\it i.e.}}
\def\eg{{\it e.g.}}
\def\no{\noindent}
\def\o{\over}

\def\bm{{\bf m}}


\def\IR{\relax{\rm I\kern-.18em R}}
\font\cmss=cmss10 \font\cmsss=cmss10 at 7pt
\def\ZZ{\relax\ifmmode\mathchoice
{\hbox{\cmss Z\kern-.4em Z}}{\hbox{\cmss Z\kern-.4em Z}}
{\lower.9pt\hbox{\cmsss Z\kern-.4em Z}}
{\lower1.2pt\hbox{\cmsss Z\kern-.4em Z}}\else{\cmss Z\kern-.4em Z}\fi}
\def\inbar{\,\vrule height1.5ex width.4pt depth0pt}
\def\IC{\relax\hbox{$\inbar\kern-.3em{\rm C}$}}
\def\IN{\relax{\rm I\kern-.18em N}}
\def\IQ{\relax\hbox{$\inbar\kern-.3em{\rm Q}$}}
\def\sL{^{(L)}}

\def\sLo{^{(L-1)}}

\def\atd{\atopwithdelims[]}


\nref\rBPZ{A.A. Belavin, A.M. Polyakov and A.B. Zamolodchikov,
 Nucl. Phys. B241 (1984) 333.}
\nref\rCFTrev{P. Ginsparg, in: {\it Fields, strings, and critical
 phenomena}, Les Houches 1988, ed.~E. Br\'ezin and J. Zinn-Justin
 (North Holland, Amsterdam, 1989).}
\nref\rKac{V.G. Kac, {\it  Infinite dimensional Lie algebras}, third
 edition (Cambridge University Press, 1990).}
\nref\rBaxbook{R.J. Baxter, {\it Exactly solved models in statistical
 mechanics} (Academic Press, London, 1982).}
\nref\rKedMc{R.~Kedem and B.M.~McCoy, J. Stat. Phys. 71 (1993) 865
 ~(hep-th/9210129);
 S. Dasmahapatra, R. Kedem, B.M. McCoy and E. Melzer,
 Stony Brook preprint, hep-th/9304150, J. Stat. Phys. (in press).}
\nref\rKKMM{R.~Kedem, T.R.~Klassen, B.M.~McCoy and E.~Melzer,
 Phys. Lett. B304 (1993) 263 and B307 (1993) 68 ~(hep-th/9211102
 and 9301046).}
\nref\rTKNS{M.~Terhoeven, Bonn preprint, hep-th/9111120; ~
 A. Kuniba, T. Nakanishi, J. Suzuki, Mod. Phys. Lett. A8 (1993) 1649
 ~(hep-th/9301018).}
\nref\rKMM{R.~Kedem, B.M.~McCoy and E.~Melzer,
 Stony Brook preprint, hep-th/9304056,
 to appear in C.N.~Yang's 70th birthday Festschrift, ed. S.T. Yau.}
\nref\rM{E. Melzer, Stony Brook preprint, hep-th/9305114,
 Int. J. Mod. Phys. A (in press).}
\nref\rABF{G.E. Andrews, R.J. Baxter and P.J. Forrester, J. Stat.
 Phys. 35 (1984) 193.}
\nref\rRoCa{A. Rocha-Caridi, in:  {\it Vertex operators in mathematics
 and physics}, ed. J.~Lepowsky, S. Mandelstam and I.M. Singer
 (Springer, Berlin, 1985).}
\nref\rAffleck{I. Affleck, Phys. Rev. Lett. 55 (1985) 1355.}
\nref\rBethe{H. Bethe, Z. Phys. 71 (1931) 205.}
\nref\rAndrbook{G.E. Andrews, {\it The theory of partitions}
  (Addison-Wesley, London, 1976).}
\nref\rTak{M. Takahashi, Progr. Theor. Phys. 46 (1971) 401.}
\nref\rFT{L.D. Faddeev and L.A. Takhtajan, Phys. Lett. A85 (1981) 375.}
\nref\rFL{A.J. Feingold and J. Lepowsky, Adv. in Math. 29 (1978) 271.}
\nref\rAGSZ{I. Affleck, D. Gepner, H.J. Schulz and T. Ziman,
 J. Phys. A22 (1989) 511.}
\nref\rFeig{B.L. Feigin and A.V. Stoyanovsky, preprint, hep-th/9308079.}
\nref\rPasq{V. Pasquier, J. Phys. A20 (1987) L1229.}
\nref\rKy{E. Date, M. Jimbo, A. Kuniba, T. Miwa and M. Okado,
 Lett. Math. Phys. 17 (1989) 51.}
\nref\rKya{M. Jimbo, K.C. Misra, T. Miwa and M. Okado,
 Commun. Math. Phys. 136 (1991) 543.}
\nref\rKir{A.N. Kirillov, private communication.}

\Title{\vbox{\baselineskip12pt\hbox{TAUP 2125-93}
\hbox{hep-th/9312043} } }
{\vbox{\centerline{The Many Faces of a Character}}}

\centerline{Ezer Melzer}

\medskip \medskip
{\vbox{\baselineskip16pt{\centerline{{\it  School of Physics and Astronomy}}
\smallskip\centerline{\it Beverly and Raymond Sackler Faculty
  of Exact Sciences}
\centerline{{\it Tel-Aviv University}}
\centerline{{\it Tel-Aviv 69978, ISRAEL}} } }
\medskip
\centerline{email: melzer@ccsg.tau.ac.il}

\vskip 10mm

\centerline{\bf Abstract}
\bigskip

We prove an identity between three infinite families of polynomials
which are defined in terms of `bosonic', `fermionic', and
`one-dimensional configuration' sums. In the limit where the
polynomials become infinite series, they give different-looking
expressions for the characters of the two integrable representations
of the affine $su(2)$ algebra at level one. We conjecture yet another
fermionic sum representation for the polynomials which is constructed
directly from the Bethe-Ansatz solution of the Heisenberg spin chain.

\Date{}

\vfill\eject

\newsec{Introduction}

General symmetry arguments lead to the observation that
certain infinite-dimensional algebras, called {\it chiral} or
{\it vertex operator} algebras, play a major role
in two-dimensional \cft~\rBPZ\rCFTrev.  In particular, the Hilbert space
of a \cft\ decomposes into irreducible highest-weight
representations of such algebras. The theory of such representations
is well developed~\rKac, and their characters, which encode
the spectrum of (certain sectors of) the \cft, have been calculated
extensively by various methods. The characters are constructed
as formal power series $\chi(q)$ in some variable $q$
(occasionally other variables -- ``counting'' the charges with
respect to some symmetry -- are also present).

Such characters also arise in studies of integrable models of
two-dimensional classical statistical mechanics and their
related (through a transfer matrix) one-dimensional quantum
systems. This comes  as no surprise
when the two- (one-)dimensional system involved is
critical (gapless), as -- having in mind the discussion
above -- scaling limits of many such systems are
expected to be described by conformal field theories.
More surprising is the appearance of characters
in computations of order parameters in {\it off}-critical
two-dimensional systems, using Baxter's corner transfer matrix
(CTM) technique~\rBaxbook.

An interesting aspect for us here is
the possibility, due to the fact that
the statistical mechanics models are  defined
 as finite systems of size $L=1,2,3,\ldots$, to
construct infinite families of polynomials
$\chi\sL(q)$ which approach the characters in the thermodynamic
limit $L$$\to$$\infty$.
In a sense, this procedure describes a gradual build up of
the infinite representation space of the relevant chiral
algebra. Now, within different frameworks one
obtains different-looking expressions for the
{\it finitized characters} $\chi\sL(q)$.
Analyses~\rKedMc~of the spectrum of
certain gapless spin-chain hamiltonians
using Bethe-Ansatz-type methods suggests that such methods
generically  lead to {\it fermionic sum}
expressions~\rKKMM-\rKMM~for the
(finitized) characters, expressions
which generalize the (finitized)
sum-side of the Rogers-Ramanujan-Schur
identities; CTM methods, on the other hand, lead naturally
to {\it one-dimensional} (1D) {\it configuration sums},
which are statistical sums
over certain restricted configurations (with appropriate weights)
defined  on a finite one-dimensional lattice.

In~\rM~we noted the equality of certain fermionic and 1D
configuration sums, which are finitized characters of the unitary
minimal models of \cft.
These 1D configuration sums appeared originally
in the work~\rABF~on RSOS models, and were shown there to be equal to
certain {\it bosonic sums}
which provide finitizations of the more familiar
expressions for the characters~\rRoCa~given in terms of
 theta functions.

In the present paper we discuss
analogous identities between bosonic, fermionic, and 1D configuration
sum expressions for finitized characters of the  $su(2)$ affine Kac-Moody
algebra at level one. The relevant spin chain in this case
is~\rAffleck~the (antiferromagnetic) spin-$\half$ Heisenberg chain, to
which the Bethe Ansatz was originally applied~\rBethe.
We will see that the Bethe-Ansatz description of its spectrum leads
naturally to a decomposition of the affine $su(2)$ characters
into characters of the so-called degenerate representations of
the Virasoro algebra at central charge $c$=1, thus motivating us to
conjecture
fermionic sum expressions also for finitizations of the latter characters.

\newsec{Polynomial Identities}

We start by recalling some standard notation (see \eg~\rAndrbook ). Let
\eqn\qsub{(q)_0=1~~,~~~~~~~  (q)_{m}=\prod_{\ell=1}^m (1-q^\ell)
  ~~~~~~{\rm for}~~m=1,2,3,\ldots,}
in terms of which the $q$-binomial coefficients are defined as
\eqn\qbin{ {n \atopwithdelims[] m}_q   ~=~
 \cases{ ~{(q)_n \o (q)_m (q)_{n-m}}
  ~~~~~~~~& if ~~$0\leq m \leq n$ \cr  ~0 & otherwise~,\cr} }
for $m,n\in \ZZ$. Note their properties~\rAndrbook
\eqn\qbinrec{ \eqalign{ {n \atd m}_q
 ~&=~ {n-1 \atd m-1}_q + q^{m}{n-1 \atd m}_q \cr
 ~&=~ {n-1 \atd m}_q + q^{n-m}{n-1 \atd m-1}_q ~~,\cr}}
\eqn\ident{ {n_1 + n_2 \atd m'}_q = \sum_{m_2\in \ZZ} q^{m_2(m_2+m'-n_2)}
  {n_1 \atd m_2+m'-n_2}_q {n_2 \atd m_2}_q~~~~~(n_1,n_2\geq 0), }
\eqn\infqbin{ \lim_{n\to \infty}{n \atd m}_q~=~{1\o (q)_m}~~~~~~~~(m\geq 0),
 ~~~~~~ \lim_{n,m\to \infty}{n+m \atd m}_q~=~{1\o (q)_\infty}~~,}
\eqn\qbinone{ \lim_{q\to 1}{n \atd m}_q~=~\pmatrix{n\cr m\cr}~=~
     {n!\o m! ~(n-m)!}~~~~~~~~(0\leq m \leq n).}

\medskip
Now for $j=0,\half$ and $L$ a non-negative integer, define the bosonic sums
\eqn\bjl{ B_j\sL(z,q) ~=~ \sum_{k\in \ZZ} z^k q^{k(k+2j)}
  {L \atd [{L\o 2}+j]+k}_q ~~,}
where $[x]$ denotes the integer part of $x$. Also, define the fermionic sums
\eqn\fjl{ F_j\sL(z,q) ~=~ \sum_{m_1,m_2\in \ZZ} z^{m_1-m_2}
   q^{{1\o 2} \bm C_{A_2} \bm^t + 2j(m_1-m_2)}
  {[{L+1\o 2}-j] \atd m_1}_q {[{L\o 2}+j] \atd m_2}_q~~,}
where $C_{A_2}=\pmatrix{2 & -1\cr -1 & 2\cr}$ is the Cartan matrix of
the algebra $A_2$, and $\bm=(m_1,m_2)$.  Finally, let
$\mu_\ell = \half(1-(-1)^\ell)$, \ie~ $(\mu_1,\mu_2,\mu_3,\ldots) =
(1,0,1,\ldots)$, and for positive $L$ (still $j=0,\half$) define the 1D
configuration sums
\eqn\cjl{ \eqalign{ C_j\sL(z,q) ~=~ (z^2 q)^{2j(-1)^L [{L+1 \o 2}]}
 & \sum_{h_1,\ldots,h_L\in \{0,1\}}  z^{\sum_{\ell=1}^L \ell
   ((h_{\ell+1}-h_\ell) - (\mu_{\ell+1} -\mu_\ell)) } \cr
  & ~\times ~  q^{ \sum_{\ell=1}^L \ell
   (\max(1-h_{\ell},h_{\ell+1})- \max(1-\mu_{\ell},\mu_{\ell+1})) }~, \cr}}
where ~$h_{L+1} = \mu_{L+1}$~ for ~$j$=0~ and ~$h_{L+1}=1$$-$$\mu_{L+1}$~
for ~$j$=$\half$.
More explicitly, since ~$\max(1-h_1,h_2)=1-h_1+h_1 h_2$ ~for~
$h_1,h_2\in\{0,1\}$, we have
\eqn\cjla{ \eqalign{ C_j\sL(z,q) ~=~ (z^2 q)^{2j(-1)^L [{L+1 \o 2}]}
 & \sum_{{h_1,\ldots,h_L\in \{0,1\} \atop h_{L+1}=(1+(-1)^{L+2j})/2}}
   z^{\sum_{\ell=1}^L \ell (h_{\ell+1}-h_\ell - (-1)^\ell)} \cr
 & ~~~~~~~~~\times ~ q^{ \sum_{\ell=1}^L \ell
   ( {1\o 2}(1-(-1)^\ell)-h_\ell(1-h_{\ell+1}) )}~, \cr} }
and in addition, for $L$=0, we set  $C_j^{(0)}(z,q)=1$.

The objects $B_j\sL(z,q)$, $F_j\sL(z,q)$, and $C_j\sL(z,q)$ are
polynomials in $q$, $z$ and $z^{-1}$ whose constant term is 1,
\ie~they can be expanded as finite (for finite $L$) sums of the
form $\sum_{n=0}^{n_1} \sum_{m=m_0}^{m_1} a_{nm} z^m q^n$
with $m_0$ a (generically) negative integer
and $a_{00}=1$ for all $L$ and both values of $j$.
For $B_j\sL(z,q)$ and $F_j\sL(z,q)$ this fact is clear from the
definitions, while for $C_j\sL(z,q)$ it is best seen from the following
theorem which is one of our main results.

\medskip \no
{\bf Theorem}: For ~$j=0,\half$ ~and ~$L=0,1,2,\ldots$
\eqn\thm{ B_j\sL(z,q)~=~F_j\sL(z,q)~=~C_j\sL(z,q) ~~.}

\no {\bf Proof}:
(i) The first equality in \thm\ is proven directly
by using the general identity \ident\
specialized to the case of $n_1=[{L+1\o 2}-j]$, $n_2=[{L\o 2}+j]$
{}~(so that $n_1+n_2=L$ ~for both ~$j=0,\half$),
and $m'=[{L\o 2}+j]+k$. The rhs of \bjl\ is then rewritten as
\eqn\prfbf{ B_j\sL(z,q)~=~
  \sum_{k,m_2\in \ZZ} z^{k} q^{k(k+2j)+m_2(m_2+k)}
   {[{L+1\o 2}-j] \atd  m_2+k}_q   {[{L\o 2}+j] \atd m_2}_q ~~,}
and the change of summation variable $k \mapsto m_1=k+m_2$ gives the
rhs of \fjl, as claimed.

\no (ii) To prove the second equality in \thm\ we show that the
$B_j\sL$ and $C_j\sL$ satisfy the same recursion relations
and initial conditions in $L$.
Starting from the rhs of \bjl, we use the first (second)
line of \qbinrec, followed by a change of summation variable
$k\mapsto k'=k+2j$ ~($k\mapsto k'=k-1+2j$)~ when
{}~${L\o 2}+j\in\ZZ$ ~(${L+1\o 2}+j\in\ZZ$)
in the second resulting sum. Thus we obtain for $L$=1,2,3,$\ldots$
\eqn\bjlrec{ \pmatrix{ B_0\sL(z,q) \cr B_{1/2}\sL(z,q)\cr} ~=~
   M\sL(z,q) \pmatrix{ B_0\sLo(z,q) \cr B_{1/2}\sLo(z,q)\cr} ~~,}
where
\eqn\MLzq{ \eqalign{ M\sL(z,q) ~&=~
  \pmatrix{1 & q^{L/2} \cr q^{L/2} & 1\cr}
    ~~~~~~~~~~~~~~~~~~~~~~~~~~{\rm if}~L~{\rm is~even} \cr
 &=~ \pmatrix{1 & zq^{(L+1)/2} \cr z^{-1}q^{(L-1)/2} & 1\cr}
     ~~~~~~~~~~{\rm if}~L~{\rm is~odd}. \cr } }
Together with the initial
conditions $B_j^{(0)}(z,q)=1$, these recursion relations
 fully determine the $B_j\sL(z,q)$ for all
$L$.  On the other hand, the recursion relations satisfied by the $C_j\sL$
are obtained by explicitly performing the summation over the variable
$h_L$ on the rhs of \cjla. Omitting the (elementary) details, we find this way
that the $C_j\sL(z,q)$ are determined by precisely the same recursion
relations and initial conditions as the $B_j\sL(z,q)$, thus completing
the proof.

\medskip
To conclude this section, let us note
that the polynomials of theorem \thm\ are
reciprocal to the Rogers-Szeg\"o polynomials (see~\eg~\rAndrbook),
\eqn\RoSz{ H_n(z;q) ~=~ \sum_{k=0}^n z^k {n \atd k}_q ~~,}
in the following sense:
\eqn\recip{ B_j\sL(z^{-1},q^{-1}) =\cases{
  z^{-[{L\o 2}]} q^{-[{L\o 2}] [{L+1\o 2}]} H_L(z;q)
      &~~if~~$L\equiv 2j$~(mod~2) \cr
  z^{-[{L+1\o 2}]} q^{-[{L+1\o 2}] [{L+2\o 2}]} H_L(zq;q)
      &~~if~~$L\equiv 2j+1$~(mod~2). \cr} }
(The derivation of \recip, starting from the definition \bjl, is elementary.)

\newsec{Connection with the Heisenberg magnet}

Here we describe how the polynomials of the previous section arise
(modulo a conjecture) in the one-dimensional
Heisenberg model, also called the spin-$\half$ XXX chain. The hamiltonian
of this model is given by the hermitean operator
\eqn\xxx{ H ~=~ J \sum_{n=1}^L ({\bf S}_n \cdot {\bf S}_{n+1}-{1\o 4})~
  ~~~~~~~~~({\bf S}_{L+1} = {\bf S}_{1}~, ~J\in \IR_{\neq 0})~,}
acting on ~$(\IC^2)^{\otimes L}$,
where the components of $2{\bf S}_n$ are the three Pauli matrices acting
on the space $\IC^2$ associated with the $n$-th site along the chain.
We will summarize only the details of the solution of the model
which are relevant for us here, following~\rTak, and refer the reader
to this reference for a more thorough discussion which also explains
the terminology used below.

The diagonalization of the hamiltonian, using Bethe's Ansatz~\rBethe,
leads to the following characterization of its eigenstates.
First of all, one exploits the $SU(2)$ symmetry of the model,
\ie~the fact that $H$ commutes with the total spin operator
{}~${\bf S}=\sum_{n=1}^L {\bf S}_n$, to diagonalize it simultaneously
with ${\bf S}^2$ (whose eigenvalues are $S(S+1)$) and $S^z$.
Thus the spectrum splits into $su(2)$ multiplets, and
each $su(2)$ highest-weight eigenstate with
$M$ down-spins and $L$$-$$M$ up-spins (where $M=0,1,\ldots,[{L\o 2}]$),
namely such that $S=S^z=\half(L-2M)\geq 0$,
is labeled by a set of
half-integers $\{\{I^a_j\}_{j=1}^{m_a}\}_{a=1}^\infty$,
where the $m_a$ satisfy
\eqn\tots{ \sum_{a=1}^\infty a m_a ~=~ M ~=~ {L\o 2} - S~.}
(The non-negative integers $m_a$ are the
number of ``strings of length $a$'', $a=1,2,3,\ldots$,
in the given state.)
For a given $a$, the $I^a_j$ are distinct
integers (half-odd-integers) if $L-m_a$ is odd  (even),
lying in ranges which depend
on the $m_a$ and the size of the system $L$:
\eqn\Irest{ |I^a_j| ~\leq~ \half \Bigl( L-\sum_{b=1}^\infty t_{ab} m_b
     -1\Bigr)~~,}
where ~$t_{ab}=2\min(a,b)-\delta_{ab}$. Note that~
$\min(a,b)=(C_{T_n}^{-1})_{ab}$~ for all $a,b\leq n$,
where $C_{T_n}$ is the Cartan matrix
of the tadpole graph of $n$ nodes, \ie~the same as the Cartan matrix of
the algebra $A_n$ except for the last entry along the diagonal
which is $(C_{T_n})_{nn}=1$.
Hence the fact that $m_a$=0 for all $a > M={L\o 2}-S$, implied by
\tots, allows us to rewrite \Irest\ as
\eqn\Iresta{ |I^a_j| ~\leq~ \half \left(L-
  (\bm(2C_{T_M}^{-1}-{\sl 1}))_a  -1\right)~~~~~~~~~~~(a=1,2,\ldots,M),}
where $\bm=(m_1,m_2,\ldots,m_M)$ and
${\sl 1}$ is the $M$$\times$$M$ identity matrix.

Now the hamiltonian \xxx\ is translational invariant, and the eigenstate
labeled by $\{I^a_j\}$ is also an eigenstate of the momentum (shift)
operator with the eigenvalue
\eqn\totp{ P ~\equiv ~ {2\pi\o L}~ \sum_{a=1}^\infty ~\sum_{j=1}^{m_a}
   \left( I^a_j + {L\o 2} \right)~~~({\rm mod}~2\pi)~.}
This formula can be interpreted as saying that
the $j$-th $a$-string in the state contributes
\eqn\pa{ p^a_j ~=~  {2\pi\o L}  \left( I^a_j + {L\o 2} \right) ~}
to the total momentum $P$.
(In fact, $p^a_j$ can be seen to be the sum of the $a$
``quasi-momenta''~\rTak~comprising the $j$-th $a$-string,
provided -- by convention -- that the real parts of the quasi-momenta
lie in the Brillouin zone $[0,2\pi]$.)
It follows from \Iresta\ that the momenta $p^a_j$
belong to the set
$\{ p^a_{\rm min}(\bm), ~p^a_{\rm min}(\bm)+{2\pi\o L},
{}~p^a_{\rm min}(\bm)+{4\pi\o L}, \ldots,~p^a_{\rm max}(\bm)\}$, where
\eqn\pmin{ p^a_{\rm min}(\bm)~=~ {\pi\o L} \left(
 (\bm(2C_{T_M}^{-1}-{\sl 1}))_a +1\right)~=~ 2\pi-p^a_{\rm max}(\bm)~~,}
and for all $a$ obey a fermionic exclusion rule
\eqn\fermi{ p^a_j ~\neq~ p^a_k~~~~~~~{\rm for}~~~~~j\neq k~.}

Let us now form the generating function of the total momentum $P$
in the sector ~$S=S^z={L\o 2}-M$,
ignoring the ~mod~$2\pi$~ in \totp, namely consider (for
$L$ a positive integer and $S\in\{0,\half,1,\ldots,{L\o 2}\}$
such that $2S\equiv L$~(mod 2)) the sum
\eqn\fsldef{ {\cal F}_S\sL(q) ~=~ q^{-M} ~\sum_{\{p^a_j\}}~
   q^{(L/2\pi) \sum_{a=1}^\infty \sum_{j=1}^{m_a} p^a_j}~~.}
The summation here is performed over all sets $\{p^a_j\}$ allowed
by \pmin-\fermi\ and \tots, and the prefactor $q^{-M}$ was introduced
for later convenience.
Using the methods developed in~\rKedMc~this sum
is recast in the fermionic form
\eqn\fsl{ {\cal F}_S\sL(q) ~=~ q^{-M} \sum_{{m_a\in\ZZ \atop
  \Sigma_{a=1}^M a m_a=M}} q^{\bm C_{T_M}^{-1} \bm^t}
  ~\prod_{b=1}^M {L-(\bm(2 C_{T_M}^{-1} -{\sl 1}))_b \atd m_b}_q~~,}
where $M={L\o 2}-S$. When $M$=0, eq.~\fsl\ should be understood
as ~${\cal F}_{L/2}\sL(q) = 1$.

\no
{\bf Conjecture}: For all ~$L\in\ZZ$ ~and ~$S\in {L\o 2}+\ZZ$
\eqn\conj{ {\cal F}_S\sL(q) ~=~ {\cal B}_S\sL(q) ~\equiv~
   {L \atd {L\o 2}-S}_q - q^{2S+1}   {L \atd {L\o 2}-S-1}_q ~~.}
Note that the rhs here serves as the definition of
${\cal B}_S\sL(q)$.

We have verified that \conj\ holds for many small values of $S$
and $L$; its validity at $q$=1, where the rhs reduces
to $\pmatrix{L\cr {L\o 2}-S\cr}- \pmatrix{L\cr {L\o 2}-S-1\cr}$
(cf.~\qbinone) which correctly counts all $su(2)$ highest-weight
states of spin $S$ in the spin chain, was proved in
appendix A of~\rTak.

The significance of the conjecture, in connection with the
discussion of sect.~2, is revealed by noting the following relation.
Starting from \bjl\ we have for $L\equiv 2j$~(mod 2)
\eqn\bjla{ \eqalign{ (z^2 q)^{j/2} B_j\sL(z,q) ~&=~
  \sum_{k\in j+\ZZ} z^k q^{k^2} {L \atd {L\o 2}+k}_q  \cr
 &=~  \delta_{j,0} {L \atd {L\o 2}}_q + \sum_{S\in j+\ZZ_{\geq 0}}
   (z^S + z^{-S}) ~q^{S^2} {L \atd {L\o 2}-S}_q  \cr
 &=~  \delta_{j,0} {L \atd {L\o 2}}_q + \sum_{S\in j+\ZZ_{\geq 0}}
   ([2S+1]_z-[2S-1]_z)~ q^{S^2} {L \atd {L\o 2}-S}_q  \cr
 &=~  \sum_{S\in j+\ZZ_{\geq 0}}
    [2S+1]_z ~q^{S^2} {\cal B}_S\sL(q) ~~,\cr} }
where $[x]_z = (z^{x/2}-z^{-x/2})/(z^{1/2}-z^{-1/2})$ ~(so that
$[2S+1]_z$ is the character ~Tr~$z^{S^z}$ of the spin-$S$
unitary representation
of the algebra $su(2)$).

It is now natural to extend the generating functions \fsldef\
to the full spectrum of the hamiltonian, consisting of whole
$su(2)$ multiplets, by forming the fermionic sums
\eqn\fjlhat{\eqalign{  \hat{F}_j\sL(z,q) ~&= ~
 (z^2 q)^{-j/2}  \sum_{S\in j+\ZZ_{\geq 0}}
    [2S+1]_z ~q^{S^2} {\cal F}_S\sL(q) \cr   &=~
 (z^2 q)^{-j/2}\sum_{S=j}^{L/2}~ [2S+1]_z~ q^{S(S+1)-L/2} \cr
  &~~\times \sum_{{m_a\in \ZZ \atop
   \Sigma_{a=1}^\infty a m_a=L/2-S}} q^{\bm C_{T_\infty}^{-1} \bm^t}
  ~\prod_{b=1}^\infty {L-(\bm(2 C_{T_\infty}^{-1} -{\sl 1}))_b
      \atd m_b}_q~~,\cr}}
where $j$=0 or $\half$, $L\equiv 2j$ (mod $2$), and the sum over
$S$ is in steps of 1.
(Here $C_{T_\infty}^{-1}$ is to
be understood as the infinite matrix whose elements are ~$\min(a,b)$;
all the sums and products in \fjlhat\ are nevertheless finite,
as automatically enforced by the restriction \tots.)
Assuming \conj\ holds, we conclude from \bjla\ that the sums
$\hat{F}_j\sL(z,q)$
provide another representation for the polynomials of theorem \thm.
The evaluation of these polynomials at $z$=$q$=1 is most easily
performed using the representation \cjl, which immediately gives
{}~$C_j\sL(1,1)=2^L$. {}From the point of view
of the spin-$\half$ Heisenberg chain the fact that
{}~$\hat{F}_j\sL(1,1)=2^L$~ demonstrates~\rBethe\rTak\rFT~the
completeness of the Bethe-Ansatz eigenstates -- taking into
account also the $SU(2)$ symmetry of the model -- for a finite chain
of $L$ sites.

\medskip

We conclude this section by making two comments. First, it is
possible to convert the expressions in eqs.~\fsl\ and \fjlhat\
into a form which suggests a relation to the fermionic sum
representations~\rKKMM\rM~for the (finitized) characters
of the unitary minimal models. To do so we make the variable
change ~$m_a \mapsto m_a' = L-2\sum_{b=1}^M m_b (C_{T_M}^{-1})_{b,M+1-a}$
{}~in \fsl, which leads to
\eqn\fsln{ \eqalign{
 {\cal F}_S\sL(q) ~=& ~q^{{L\o 2}({L\o 2}-1)-S(S-1)}
  \sum_{m_2',\ldots,m_M' \in 2(\ZZ+S)}
    q^{{1\o 4}\bm' C_{A_M} \bm'^t -{1\o 2}L m_M'}  \cr
  &~~~~~~~\times ~\prod_{b=1}^M
  {{1\o 2}((\bm'I_{A_M})_b +m_1' \delta_{b,1}+L \delta_{b,M})
    \atd m_b'}_q~~,\cr} }
where $M={L\o 2}-S$ as before, $m_1'=2S$ (which follows from \tots),
and $C_{A_M}$ ($I_{A_M}$) is the
Cartan (incidence) matrix of the algebra $A_M$.
Similarly, eq.~\fjlhat\ can be rewritten as
\eqn\fjlhatn{ \eqalign{ &\hat{F}_j\sL(z,q) ~=~ (z^2 q)^{-j/2}
   q^{{L\o 2}({L\o 2}-1)}
  \sum_{m_1',\ldots,m_M' \in 2(\ZZ+j)} [m_1'+1]_z  \cr
  &~~\times~ q^{{1\o 4}\bm' C_{A_M} \bm'^t +{1\o 2}(m_1'-L m_M')}
 \prod_{b=1}^M
  {{1\o 2}((\bm'I_{A_M})_b +m_1' \delta_{b,1}+L \delta_{b,M})
    \atd m_b'}_q~~,\cr} }
where $M=(L-m_1')/2$.

Second, recall that the sum \fjlhat\ originated from the quasiparticle sum
\fsldef\ which -- potentially, as $L$$\to$$\infty$ -- involves
infinitely many types of quasiparticles $a$=1,2,3,$\ldots$.
We contrast it here with the 2-quasiparticle
sum \fjl, explicitly exhibiting the latter's quasiparticle interpretation.
As in~\rKedMc, we can rewrite \fjl\ as
\eqn\fjlqp{ F_j\sL(z,q) ~=~ \sum_{m_\pm =0}^\infty z^{m_+ -m_-}
  \sum_{\{p^\pm_j\}_{j=1}^{m_\pm}}
   q^{(L/2\pi) \sum_{a=\pm} \sum_{j=1}^{m_a} p^a_j}~~,}
where ~$p_j^a \in \{ p^a_{\rm min}(m_\pm),~p^a_{\rm min}(m_\pm)+{2\pi\o L},~
\ldots,~p^a_{\rm max}(m_\pm)\}$,  ~$p_j^a\neq p_k^a$~ for ~$j\neq k$,
and
\eqn\twoqpr{ \eqalign{ p_{{\rm min}}^+(m_\pm)~&=~
  {\pi\o L} (m_+ -m_- +4j+1)~=~p_{{\rm max}}^+(m_\pm)
   -{2\pi\o L}[{\textstyle {L-1\o 2}}-j]  \cr
 p_{{\rm min}}^-(m_\pm)~&=~
  {\pi\o L} (m_- -m_+ -4j+1)~=~p_{{\rm max}}^-(m_\pm)
   -{2\pi\o L}([{\textstyle {L\o 2}}+j]-1)~~.  \cr}}

\newsec{Characters}

In the limit $L$$\to$$\infty$
the polynomials discussed in the previous sections become
infinite series. We will use the symbols $C_j^{(\infty)}(z,q)$,
${\cal F}_S^{(\infty)}(q)$, and $\hat{F}_j^{(\infty)}(z,q)$
to denote the $L$$\to$$\infty$ limits of the rhs's of eqs.~\cjla,
\fsl, and \fjlhat, respectively. In the case of the
$B_j\sL(z,q)$, $F_j\sL(z,q)$, and ${\cal B}_S\sL(q)$ the
following simplifications occur due to \infqbin:
\eqn\bosch{ B_j^{(\infty)}(z,q)~=~{1\o (q)_\infty} ~\sum_{k\in\ZZ}
     z^k q^{k(k+2j)}~~~,~~~~~~~~
   {\cal B}_S^{(\infty)}(q)~=~{1-q^{2S+1}\o (q)_\infty}~~,}
\eqn\ferch{ F_j^{(\infty)}(z,q) ~=~ \sum_{m_1,m_2=0}^\infty z^{m_1-m_2}~
   {q^{{1\o 2} \bm C_{A_2} \bm^t + 2j(m_1-m_2)} \o (q)_{m_1} (q)_{m_2}}~~.}

The series in \bosch\ are familiar
expressions for characters of two important
infinite-dimensional Lie algebras:
$B_j^{(\infty)}(z,q)$ with $j=0,\half$ are~\rKac\rFL~the
(normalized) characters ~$\chi_j = {\rm Tr}~
z^{J^z_0-j} q^{L_0-j/4}$~  of the two integrable highest-weight
representations of $(A_1^{(1)})_1$, the $su(2)$ affine Kac-Moody
algebra at level 1; the ${\cal B}_S^{(\infty)}(q)$, on the other
hand, are (see \eg~\rCFTrev)
the characters ~$\chi^{{\rm Vir}}_S = {\rm Tr}~q^{L_0-S^2}$~
of the irreducible highest-weight representations of
the Virasoro algebra at central charge $c$=1 and highest-weight
$\Delta=S^2$, ~$S\in {1\o 2}\ZZ_{\geq 0}$. The well-known expansion
\eqn\chexp{ (z^2 q)^{j/2} \chi_j(z,q) ~=~ \sum_{S\in j+\ZZ_{\geq 0}}
   [2S+1]_z~ q^{S^2} \chi_S^{{\rm Vir}}(q)~~,}
showing the decomposition of the $(A_1^{(1)})_1$ representations
into (degenerate) Virasoro representations,
can be recovered from eq.~\bjla\ in the limit $L$$\to$$\infty$.

Thus we have

\no
{\bf Corollary} (of theorem \thm):
\eqn\coro{\chi_j(z,q)~=~ F_j^{(\infty)}(z,q)~=~ C_j^{(\infty)}(z,q)~~~
 ~~~~~~~~~(j=0,\half)~.}
In addition, validity of the conjecture \conj\ implies in the limit
$L$$\to$$\infty$
\eqn\corconj{ \chi_S^{\rm Vir}(q) ~=~ {\cal F}_S^{(\infty)}(q)~~~~
 ~~~~~~~~~(S=0,\half,1,\ldots) ~,}
and if \corconj\ is true, eq.~\chexp\ yields the identity
\eqn\coroa{\chi_j(z,q)~=~ \hat{F}_j^{(\infty)}(z,q)~~~~
 ~~~~~~~~~(j=0,\half)~.}

\medskip
The rhs of eq.~\coroa\ was constructed in sect.~3 from
the Bethe-Ansatz data of the spin-${1\o 2}$ XXX chain.
Its equality to the characters of $(A_1^{(1)})_1$ provides further
support to the observation~\rAffleck~that
the long-distance asymptotics of the
antiferromagnetic chain is described by the level one $SU(2)$ WZW model
(perturbed by some marginal operators~\rAGSZ).
The work of~\rAffleck~is concerned mainly with
the symmetry of the model in the continuum limit, while \coroa\
is a statement about the degeneracy structure of the spectrum.

\newsec{Discussion}

Several comments on eqs.~\coro-\coroa\ are in order.
The first equality in \coro\ (specialized to $z$=1) has recently appeared
in~\rFeig, where a new construction of highest-weight representations
of certain subalgebras of affine Lie algebras was presented.
In particular, representations of $A_1^{(1)}$ were obtained there
by considering a subalgebra of $A_2^{(1)}$ which is isomorphic to
$A_1^{(1)}$ (this ``explains'' the appearance of the Cartan matrix
of $A_2$ in \ferch).

The equality ~$\chi_j(z,q)= F_j^{(\infty)}(z,q)$~
can easily be inferred also from the
results of~\rKMM. In the latter work fermionic expressions of the
type \ferch\ were given for the theta functions
\eqn\theta{ f_{a,b}(z,q) ~=~ {1\o (q)_\infty}~\sum_{k\in \ZZ}
   z^{k+{b\o 2a}} q^{a(k+{b\o 2a})^2}~~~~~~~~~~~~~
   (a\in \ZZ_{>0}~, ~~b\in \ZZ).}
Taking $a$=1 and $b$=$2j$=0,1, where ~$f_{1,2j}(z,q) = (z^2 q)^{j/2}
 B_j^{(\infty)}(z,q)$,
they yield the representation
\eqn\oldfch{ \chi_j(z,q) ~=~
  \sum_{{m_1,m_2=0 \atop m_1-m_2\equiv 2j({\rm mod}~2)}}^\infty
  z^{{m_1-m_2\o 2}-j}~{q^{({m_1+m_2\o 2})^2-j^2} \o (q)_{m_1} (q)_{m_2}}~}
for the $(A_1^{(1)})_1$ characters, which is slightly different
from \ferch. However, if
the expressions given in~\rKMM~for the $f_{4,b}(z,q)$
are inserted into the rhs of the simple identity ~$f_{1,2j}(z,q) =
f_{4,4j}(z^2,q)+f_{4,4(j+1)}(z^2,q)$,
then \ferch\ is obtained.

The main difference between the 2-quasiparticle sums \ferch\ and \oldfch\
is in the quadratic form in the exponent of $q$.
In fact, there exists an infinite family of such sums
with different quadratic forms. Their finitized version is obtained
by considering the following generalization of $B_j\sL(z,q)$,
\eqn\bjln{ B_j^{(L;N)}(z,q) ~=~ \sum_{k\in \ZZ} z^k q^{k(k+2j)}
  {L \atd [{L\o 2}+j]+Nk}_q ~~~~~~(N=1,2,3,\ldots),}
which reduces to \bjl\ when $N$=1.
Performing essentially the same manipulations as in part (i) of the proof
of theorem \thm\ leads to the identity
{}~$B_j^{(L;N)}(z,q)=F_j^{(L;N)}(z,q)$, where
\eqn\fjln{\eqalign{ &F_j^{(L;N)}(z,q)  \cr
  &~~~~= \sum_{m_1,m_2\in \ZZ \atop m_1-m_2\equiv 0({\rm mod}~N)}
   z^{{m_1-m_2\o N}}
   q^{{1\o 2} \bm B_N \bm^t + {2j\o N}(m_1-m_2)}
  {[{L+1\o 2}-j] \atd m_1}_q {[{L\o 2}+j] \atd m_2}_q~~.\cr}}
Here $B_N=\pmatrix{2/N^2 & 1-2/N^2 \cr 1-2/N^2 & 2/N^2\cr}$, and so
 ~$B_1=C_{A_2}$~ while ~${1\o 2} \bm B_2 \bm^t =
({m_1+m_2\o 2})^2$, corresponding to \ferch\ and \oldfch, respectively.
Now in the infinite $L$ limit we have ~$\chi_j(z,q)=B_j^{(\infty;N)}(z,q)
= F_j^{(\infty;N)}(z,q)$~ independently of $N$. But at finite $L$ the
sums \bjln-\fjln\ do depend on $N$. In particular, their value
at $z$=$q$=1 is easily calculable when ~$L\equiv 0~({\rm mod}~2N)$:
\eqn\bjlnone{ \eqalign{ B_j^{(L;N)}(1,1)~ &=~
    \sum_{k=0 \atop k\equiv 0({\rm mod}~N)} \pmatrix{L \cr k\cr} \cr
 &= ~{1\o N}\sum_{j=0}^{N-1} \biggl(2\cos{j\pi\o N}\biggr)^L ~=
   ~{1\o 2N}~ {\rm Tr}~\bigl(I_{A_{2N-1}^{(1)}}\bigr)^L ~~,\cr}}
where $I_{A_n^{(1)}}$ is the incidence matrix of the Dynkin diagram of the
affine Lie algebra $A_n^{(1)}$.

The most rhs of eq.~\bjlnone\ is naturally
interpreted as the total number of states in a ``spin chain'' of $L$
sites with periodic boundary conditions, where the
states (or heights) ~$h_\ell\in\{0,1,\ldots,2N-1\}$~
at each site $\ell=1,2,\ldots,L$ are labeled by the nodes of the
Dynkin diagram of $A_{2N-1}^{(1)}$ whose incidence matrix imposes the
adjacency constraint ~$h_{\ell+1}-h_\ell \equiv \pm 1~({\rm mod}~2N)$,
and furthermore one of the heights is fixed (say ~$h_1$=0).
This suggests a connection between the polynomials \bjln-\fjln\ and
the spectrum of the hamiltonian obtained from the transfer matrix
of the critical RSOS lattice model based on $A_{2N-1}^{(1)}$~\rPasq.
The conformal field theory associated with the latter model is
identified~\rPasq~as the $c$=1 gaussian model at compactification
radius $r={N\o \sqrt{2}}$, which is~\rCFTrev~a $\ZZ_N$
orbifold of the level one $SU(2)$ WZW model.
This orbifold relation allows the construction of the affine
characters ~$(z^2 q)^{j/2}\chi_j(z,q) = f_{1,2j}(z,q)$~ as the sums
{}~$\sum_{b=1}^{N} f_{N^2,2(b+j)N}(z^N,q)$~ of characters of the
$r = {N\o \sqrt{2}}$ gaussian model.
The fact that the
affine characters $\chi_j$ are recovered in the $L\to\infty$ limit
of \bjln-\fjln\ is therefore consistent with the connection suggested
above.

Next, the identity ~$\chi_0(1,q)=C_0^{(\infty)}(1,q)$, which is a special
case of the second equality in~\coro, was first shown in~\rKy,
where an explicit construction of the vacuum representation
of $(A_n^{(1)})_1$ on a space of paths in $\ZZ_{n+1}$ is presented.
This construction was generalized to higher levels and arbitrary
integrable representations in~\rKya.
As far as we know, the conjectures \corconj-\coroa\ are new.

\medskip
Returning to sect.~2,
it would be interesting to generalize  theorem \thm\
to polynomial identities between different forms of
finitized characters of representations of higher rank and level
affine Lie algebras.
The relevant fermionic and 1D configuration
sum expressions for the characters themselves are known in various
cases~\rFeig-\rKya.
The connection between the finitized
characters and the spectrum of finite spin chains, discussed
in sect.~3 for the case of $(A_1^{(1)})_1$, should also be further
explored and generalized. This may improve our
understanding of the scaling limits of  spin chains which
are relevant to quantum field theory.

\vskip 18mm

\no
{\it Note added.}
I have been recently informed by A.N.~Kirillov that he has a
proof~\rKir~of conjecture \conj\ (and hence also of eq.~\corconj).

\vskip 18mm

\no
{\it Acknowledgements.}
I would like to thank A.~Berkovich, F.~E\ss ler, V.~Korepin and
B.~McCoy for useful discussions.
This work was supported in part by the NSF, grant 91-08054,
and by the US-Israel Binational Science Foundation.

\vfill\eject

\listrefs

\vfill\eject

\bye\end